\documentstyle[twocolumn,epsf,aps]{revtex}
\draft
\begin{document}
\title{Evolutionary prisoner's dilemma game on a square lattice}
\author{Gy\"orgy Szab\'o$^1$ and Csaba T\H oke$^2$}
\address
{$^1$Research Institute for Materials Science, POB 49, H-1525 Budapest,
Hungary \\
$^2$E\"otv\"os University, M\'uzeum krt. 6-8, H-1088 Budapest, Hungary}
\address{\em \today}
\address{
\centering{
\medskip \em
\begin{minipage}{14.5cm}
{}~~~A simplified prisoner's game is studied on a square lattice when the
players interacting with their neighbors can follow two strategies:
to cooperate (C) or to defect (D) unconditionally. The players updated
in random sequence have a chance to adopt one of the neighboring strategies
with a probability depending on the payoff difference. Using Monte Carlo
simulations and dynamical cluster techniques we study the density $c$ of 
cooperators in the stationary state. This system exhibits a continuous
transition between the two absorbing state when varying the value of
temptation to defect. In the limits $c \to 0$ and 1 we have observed critical
transitions belonging to the universality class of directed percolation.
\end{minipage}
}}
\maketitle

\narrowtext

\section{Introduction}
\label{sec:intr}

The evolutionary prisoner's dilemma games were introduced by Axelrod
\cite{axelrod} to study the emergence of cooperation rather than
exploitation among selfish individuals. Since the pioneering work of
Axelrod this approach has become a fruitful tool in the area of
political and behavior sciences, biology and
economics \cite{sigmund,sciam,smith}.

In the prisoner's dilemma (PD) game each of two players have to decide
simultaneously whether it wishes to cooperate with the other or to defect.
The rewards dependent on their choices are expressed by $2\times 2$ 
payoff matrices in agreement with the four possibilities. Assuming
symmetric game the players get rewards $R$ ($P$) if both choose to 
cooperate (defect). In the remaining two cases the defector's and 
cooperator's payoff are $T$ (temptation to defect) and $S$ (sucker's
payoff) respectively. The elements of payoff matrix satisfy the
following conditions: $T>R>P>S$ and $2R>T+S$. In this game the mutual
cooperation leads to the highest total (average) payoff. The highest
individual payoff ($T$) can only be reached against the other 
player decreasing the average payoff. These features makes
the PD game so interesting in the mentioned areas.

In earlier studies $N$ contestants played an iterated round-robin
prisoner's dilemma game. The population of contestants, which apply
different algorithms to choose between defection and cooperation
in the knowledge of previous decisions, was
modified according to a Darwinian selection rule round by round.
For example, eliminating the worst player the best one will have an
offspring inheriting the parent's strategy. In a different
interpretation, the worst player adopts the best algorithm. Computer
tournaments (simulations) were performed to study how the population of
contestants varies with time \cite{axelrod}. Evidently, the final (stationary)
state depends on the initial population. The simulations
have clarified the emergence of mutual cooperation among all the players
under some conditions. In these tournaments the winner the so-called 
'Tit for Tat' (TFT) algorithm has a crucial role. This very simple
algorithm cooperates in the first round and later it reciprocates the
partner's previous decision. It forces the players to 
cooperate mutually and maintains this state against defectors.

Beside the homogeneous system with players following TFT algorithm,
the state where all the players choose to defect 
has proved to be stationary too, more precisely, spare cooperators will
be suppressed due to the evolutionary rule in the large $N$ limit. 
More precisely, only a sufficiently large portion of mutual cooperators
can survive among defectors. The emergence of uniform cooperation becomes
easier when -- combining the evolutionary game with spatial effects -- the
players interact much more with their neighbors than with those who are
far away as it is typical in real populations. The spatial effects promote
the survival of cooperators even if we do not use any kind of elaborate
strategies such as the TFT.

Recently Nowak and May \cite{NM} have introduced a spatial evolutionary
PD game. In this model individuals located on a lattice
play with their neighbors and with themself. The strategical complexities
and memories of past encounters are neglected by considering only two simple
kinds of individuals: those who cooperate ($C$) and those who defect ($D$)
unconditionally. The evolutionary rule was also simplified by using discrete
time steps. Between two rounds individuals adopt the strategy
that has received the highest payoff among its neighbors including themself.
This deterministic model is equivalent to a two-state cellular automaton
where the next state at a given lattice point is determined by the
states on the surrounding points. The outcome depends on the
initial configuration and the rescaled payoff matrix described by a
single parameter $b$ characterizing the measure of temptation to defect
(see the matrix in Sec.~2). 
This model with and without self-interaction was investigated on
different lattice structures (square, triangle, cubic). The most
exhaustive analysis is performed on a square lattice taking into
account the interactions with the first and second neighbors and
self-interaction. Nowak and May observed a rich variety of spatial and temporal
dynamics dependent on the value of $b$. For example, the cooperators can
invade the word of defectors along straight border lines while defectors
gain along irregular boundaries for a given interval of $b$. Furthermore,
the above rules conserve the symmetries of the initial state for adequate
boundary conditions. Due to the discrete nature of total payoff there
appear sharp steps when varying $b$.

Introducing stochastic evolutionary rules between two rounds
Nowak {\it et al.} \cite{NBM} have extended the above model. 
Although the stochasticity simplifies the dynamics it does not change
the basic observations that cooperators and defectors can coexist.
The randomness destroys the straight border lines as well as other
symmetries that appear in the deterministic model.

Hubermann and Glance \cite{hubermann} have studied a similar model
using continuous time simulations where players are chosen randomly and
immediately updated. Their results support that the above conclusions
are not affected by whether we use continuous or discrete time
in the stochastic case \cite{NBM}.

In the present work we study a PD game with a slightly different
continuous time evolution on square lattice. In the modified model
the players need less intelligence to decide whether they adopt one of
the neighboring strategy or not. Using systematic Monte
Carlo simulations and generalized mean-field techniques we calculate
the density of cooperators as a function of $b$ for different noise
levels. It will be shown that the transitions from the active state
(coexistence of defectors and cooperators) to the absorbing ones
(all $D$ or all $C$) exhibit universal behavior.

\section{The model}
\label{sec:model}

The players located on a square lattice can follow only two simple
strategies: $C$ (always cooperate) and $D$ (always defect). Due to
this simplification this system can be handled with the Ising formalism
and we can use the sophisticated techniques developed in
non-equilibrium statistical physics. 
Each player plays a PD game with itself and with its neighbors. The total
payoff of a certain player is the sum over all interactions.
The elements of payoff matrix can be
rescaled because the evolutionary rule depends on the payoff differences
between the players. 
Accepting the idea suggested by Nowak and May \cite{NM} we choose
$R=1$, $P=S=0$ and $T=b$. Thus, the payoff to player $A$ against $B$
is given by the matrix:
\begin{center}
\begin{tabular}{|r|l|l|} \hline
$_A$ $^B$  & C & D \\ \hline
C & 1 & 0 \\ \hline
D & $b$ & 0 \\ \hline
\end{tabular}
\end{center}
where $b>1$.

Two systems will be considered subsequently. In the first case
only the first neighbors are taken into account. This means that
the total payoff of a defector surrounded by cooperators is $4b$
while the cooperator's payoff is $5$ in the same surroundings.
In the second case the neighborhood includes the first- and 
second-neighbors. Thus the payoffs of the defector and
cooperator are $8b$ and $9$ in the sea of cooperators.

The randomly chosen player $X$ revises its strategy according to the 
following rules. This player selects one of its neighbours $Y$ with
equal probability. Given the total payoffs ($E_X$ and $E_Y$) from
the previous round, player $X$ adopts the neighbor's strategy with 
the probability:
\begin{equation}
W = \frac{1}{1 + \exp{[-(E_Y - E_X)/K]}}
\end{equation}
where $E_Y$ is the neighbor's payoff and $K$ characterizes the noise
introduced to permit irrational choices. For successful strategy adoptation
the new state as well as the new payoffs are updated. Notice that the decision
is not affected by the variation of total payoff involving the change
in the surroundings. Starting from a random initial state the above
process is repeated many times.

For $K=0$ the player $X$ adopts $Y$'s strategy if $E_Y>E_X$. In this
case the randomness is represented by the selection of the players
$X$ and $Y$. The finite value of $K$ characterizes the range of payoff
difference within which the irrational decision can typically appear.
At present, our analysis is constrained to noise levels $K<1$.

Monte Carlo (MC) simulations are performed varying the value of $b$ for
fixed $K$ values. We have determined the density $c$ of cooperators using
periodic boundary conditions. The system size was varied from $L=200$
to 1000; the large sizes are required to suppress the statistical error
in the critical regions ($c \to 0$ or 1).

The above models are also investigated by the generalized mean-field technique
that proved to be very efficient for studying dynamical systems such as
the one-dimensional stochastic cellular automata \cite{gut,1DCA,OBS} and
driven lattice gases \cite{dlg1d,dickman,gmf}. In fact the introduction
of the above evolutionary rule is motivated by the demand to make the
model more convenient for this method. In the present case
we have adapted the two-dimensional method to determine
the probability of the configurations appearing on 2-, 4-, 5- and
6-point clusters \cite{gmf}. It is expected that the larger the cluster we use
the more accurate the prediction given by this technique. At the level of
6-point approximation -- taking the consistency conditions and symmetries
into account -- we have to determine 20 parameters by solving a set of
equation of motions for the configuration probabilities in the
stationary state. Details of this calculations are given in
previous papers \cite{dickman,gmf}.

\section{Results}
\label{sec:results}

For both models the $c=0$ (all $D$) and 1 (all $C$) states are independent
of time because the evolutionary rule cannot create a new strategy which
can spread away under advantageous conditions. 
The uniform cooperation ($c=1$) is a stable state if $b$ does
not exceed a threshold value $b_{c1}$ which is larger than 1. This means
that any constellation of defectors will be defeated if $b<b_{c1}$.
In the same way the $c=0$ state remains stable for $b>b_{c2}$.
Henceforth we will concentrate on those states which the cooperators
and defector can coexist in, that is, when $b_{c1}<b<b_{c2}$.

\begin{figure}
\centerline{\epsfxsize=8cm
            \epsfbox{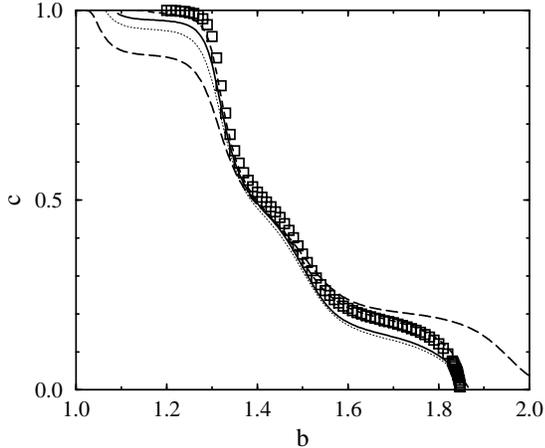}
            \vspace*{0.5mm}   }
\caption{Density of cooperators as a function of temptation to defect for
$K=0.1$. The MC data are plotted by squares, the results of generalized
mean-field technique for different cluster sizes are indicated by
long-dashed (2), dotted (4), dashed (5) and solid
(6) lines.}
\label{fig:c_b01}
\end{figure}

First we consider the model with first-neighbor interactions.
Figure~\ref{fig:c_b01} shows the $b$-dependence of the density $c$ of
cooperators in the coexistence region for $K=0.1$. As indicated
$c$ decreases monotonously with increasing $b$ until the
second threshold $b_{c2}$ where the cooperators vanish.

\begin{figure}
\centerline{\epsfxsize=7cm
            \epsfbox{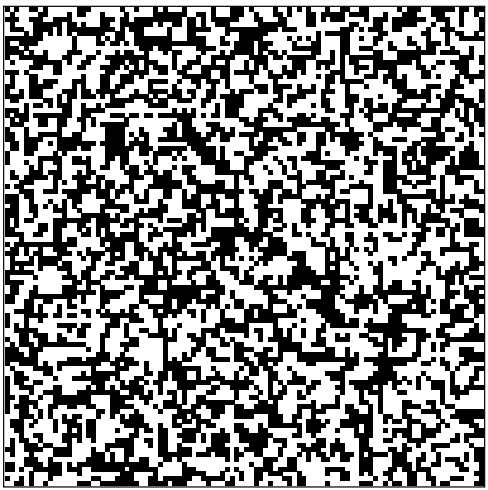}
            \vspace*{2mm}   }
\caption{Distribution of cooperators (white boxes) and defectors
(black boxes) for $b=1.4$ and $T=0.1$ ($c=0.515$).}
\label{fig:coex}
\end{figure}

The results of both the MC simulations and the generalized mean-field
method refer to step-like behavior becoming more and more striking
if we decrease the value of $K$. The sharp steps appear at the break
points (e.g. $b=4/3$, $3/2$) described by Nowak and May \cite{NM}.
Inside the coexistence region the mean-field results of 4-, 5- and
6-point approximations agree satisfactorily
with the simulations while the pair approximation yields well-marked
difference. The best agreement is found for the 5-point approximations
(dashed line).

A typical snapshot on the steadt state distribution of cooperators and
defectors is illustrated in Fig.~\ref{fig:coex} for $b=0.4$ and $T=0.1$.
This snapshot as well as the subsequent ones are a $100 \times 100$
portion of the full $400 \times 400$ lattice.
In this case the pair- (2-point) approximation gives a satisfactory
description of the short-range correlations.

Notice furthermore that the mean-field predictions are not adequate
when $c$ tends either to 0 or 1. Namely, the 4- and 6-point
approximations predict a continuous (linear) transition, the 5-point
approximation indicates a first-order one while the simulations
suggest power law behavior if $c \to 0$. Similar situation has
already been observed for a one-dimensional stochastic cellular
automaton \cite{OBS}. The mentioned deviations are not surprising
because the mean-field approximations are not capable to handle the
critical transitions exhibiting enhanced fluctuations and long-range
correlations.

\begin{figure}
\centerline{\epsfxsize=7cm
            \epsfbox{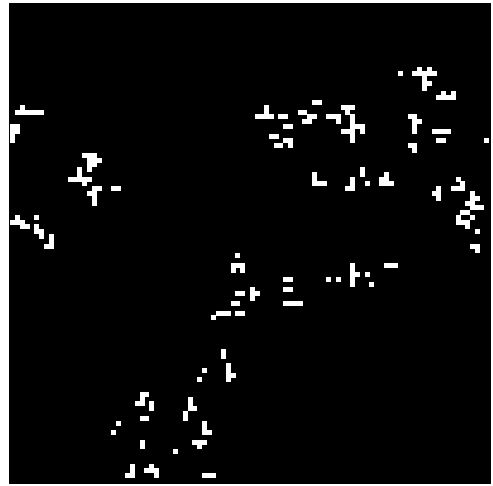}
            \vspace*{2mm}   }
\caption{The cooperators (white boxes) form colonies in the sea of
defectors (black boxes) when their density goes to 0.}
\label{fig:colon}
\end{figure}

In the limit $c \to 0$ the cooperators can survive if they form scattered
colonies in the background of defectors as illustrated in
Fig.~\ref{fig:colon}. In general, any compact
colony formation would be more preferable for cooperators,
however, the defectors make them rare. 

Visualizing the time-dependent configuration one can observe
how the colonies try to spread away. Their center, size and shape
change continuously and a separated colony can disappear without trace.
Two colonies can unite providing better opportunity for their survival,
or oppositely, a colony can divide into two (or more) parts. Similar
phenomena can be observed in a wide range of dynamical processes
described by the directed percolation (DP) \cite{DP}, the Reggeon field
theory \cite{RFT}, the surface-reaction \cite{SRM}
and Schl\"ogl models \cite{schlogl} as well as the branching and
annihilating random walks \cite{BARW}. Grassberger \cite{grass82} and
Janssen \cite{janssen} conjectured that all one-component models
with a single absorbing state belong to the universality class of
directed percolation. Exceptions can appear if the dynamics
conserves some symmetries (e.g. parity of offsprings).

Our MC data (shown in Fig.~\ref{fig:c_b01}) refer to a power law behavior,
that is
\begin{equation}
c \propto (b_{c2}-b)^{\beta}
\end{equation}
if $b \to b_{c2}$. The best fit is obtained for $b_{c2}=1.8472(1)$ and
$\beta = 0.56(3)$ which is consistent with the critical exponent
($\beta \approx 0.58$) of the two-dimensional directed percolation
\cite{exponent}.

\begin{figure}
\centerline{\epsfxsize=7cm
            \epsfbox{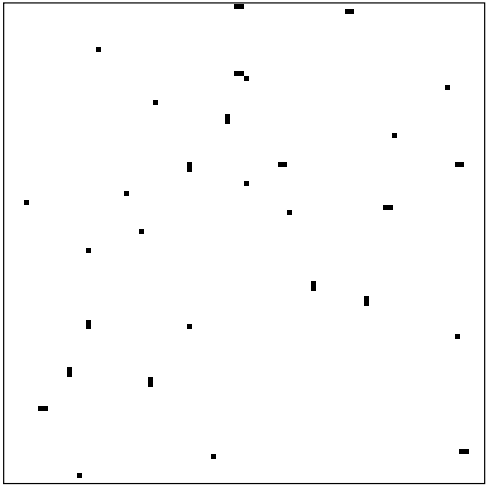}
            \vspace*{2mm}   }
\caption{Typical snapshot for high concentrat
ion of cooperators
(white boxes) for $b=1.222$ and $T=0.1$. The small 'gangs' of defectors
(black boxes) walk randomly.}
\label{fig:gangs}
\end{figure}

Contrary to the above pattern defectors form small isolated
'gangs' as demonstrated in Fig.~\ref{fig:gangs} for a typical
stationary state if $1-c \ll 1$. A single defector surrounded by
cooperators has the highest payoff (fitness) in this system.
Sooner or later this defector will have a neighboring offspring which
reduces its payoff immediately. (This process can be considered as
a retaliation executed by the TFT algorithm if more elaborate
strategies are permitted.) If $b<4/3$ then one of the defectors
will be defeated within a short time. The iteration of this process
yields randomly walking gangs. Two colliding gangs can unite into one.
Due to the possibility of irrational choices a single gang can
divide into two or can disappear. Shortly, the gangs can be considered
as branching and annihilating random walkers whose critical behavior
belongs to the DP universality class too. 

In the deterministic model introduced by Nowak and May \cite{NM}
isolated gangs with fixed positions can occur if $1<b<4/3$. The
density of gangs (whose size alternates cyclically if $5/4<b<4/3$)
depends on the initial state. In contrary to this feature the homogeneous
cooperation can emerge in the stochastic models even for $b>1$ as a
consequence of the random walk and annihilation. Besides, the
random walk causes the steady state density to be independent of
the initial state.

Despite the mentioned expectation the MC data in Fig.~\ref{fig:c_b01}
do not show any power law behavior in the limit $c \to 1$. This
discrepancy can be resolved by reminding the reader that the critical
behavior is controlled by a simple function of the diffusion constant and
the rates of branching and annihilation. In the present case these
parameters are strongly non-linear functions of $b$ at low $K$.
For higher value of $K$, however, the non-linear contributions are
negligible in the close vicinity of $b_{c1}$ and we expect the power law
behavior to appear clearly. In order to check this statement we have
repeated the same analyses at higher noise level.

\begin{figure}
\centerline{\epsfxsize=8cm
            \epsfbox{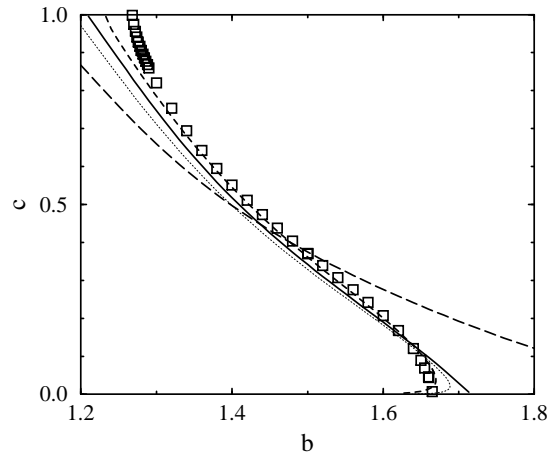}
            \vspace*{0.5mm}   }
\caption{Density of cooperators {\it vs.\ } $b$ for $K=0.5$ as suggested
by MC simulations (squares) and generalized mean-field approximations whose
level is indicated as in Fig.~1.}
\label{fig:c_b05}
\end{figure}

The results obtained for $K=0.5$ are summarized in Fig.~\ref{fig:c_b05}.
As expected the MC data show power law behavior for both the
limits $c \to 0$ and 1. Detailed numerical analysis results in
$b_{c1}=1.2687$,  $\beta = 0.62(5)$ if $c \to 0$ and $b_{c2}=1.6644(2)$,
$\beta = 0.59(3)$ if $c \to 0$. These $\beta$ values agree satisfactorily
with the corresponding exponent of DP universality class. Notice, furthermore,
that $b_{c1}$ and $b_{c2}$ depend on $K$. The determination
of a $K-b$ phase diagram indicating the active and absorbing states
goes beyond the purpose of the present work. Instead of it we have studied
the model involving second-neighbor interactions.

\begin{figure}
\centerline{\epsfxsize=8cm
            \epsfbox{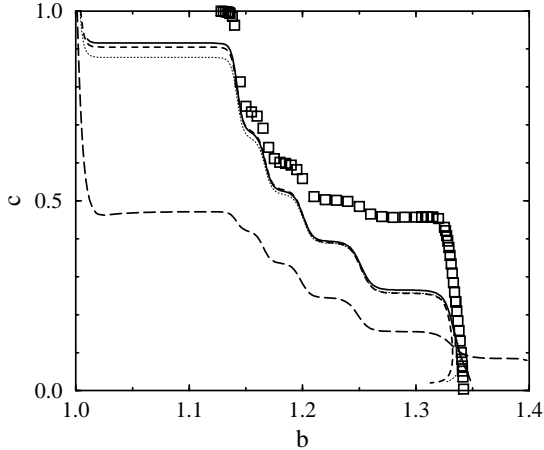}
            \vspace*{0.5mm}   }
\caption{Density of cooperators as a function of $b$ for the model
taking the second neighbor interactions into account at $K=0.02$.
Results are indicated as in Fig.~1.}
\label{fig:c_bnnn}
\end{figure}

The generalization of our techniques to investigate the density of
cooperators in the second model is straightforward. The results
of these calculations (see Fig.~\ref{fig:c_bnnn}) refer to a behavior
similar to those of the previous version. There are some minor
differences. For example, the threshold values ($b_{c1}$ and $b_{c2}$) are
definitely smaller than those of the previous model. Furthermore,
the convergence of the results of generalized mean-field approximation
is slow. This fact indicates that the short-range correlations 
become more relevant if we take the second-neighbor interactions
into account.

The steps of the continuous $c(b)$ function (for $b=8/7$, $7/6$, $6/5$
and $5/4$) becomes sharper when decreasing the value of $K$.
For high noise levels the function becomes smooth and exhibits power
law behavior with exponents close to the DP value at both ends of
the active region. Inside the active phase the difference between the
mean-field results and MC simulations decreases with increasing $K$.

\section{Conclusions}
\label{sec:conc}

We have studied the evolution of cooperation among players who can 
follow only two simple strategies ($C$ and $D$) and are placed
on a square lattice. The individual receives payoffs from
interactions with each of its neighbors and itself in a PD confrontation.
An evolutionary rule is introduced by slightly modifying the model
suggested by Nowak {\it et al} \cite{NBM}. Namely, a randomly chosen
player is to adopt one of its neighboring strategy with a
probability dependent on the payoff difference. Two versions of
the model have been investigated. In the first case the neighborhood is
limited to the first-neighbors. In the second case we have increased
the number of neighbors by taking into consideration the second-neighbors
too.

The MC simulations have given direct evidence of the existence of two
absorbing states ($c=1$ if $b<b_{c1}$ and $c=0$ if $b>b_{c2}$).
It is remarkable that the homogeneous cooperation proved to be
stable against the temptation to defect for $1<b<b_{c1}$ due to the
randomness and possibility of irrational choice. We have found
significantly different ($K$-dependent) threshold values in the models
we are interested in. It is expected that $b_{c1}$ tends to 1 if we
increase the number of neighbors.

For high density of defectors the cooperators
forming compact blocks can spread if $b<b_{c2}$. Comparing the present
models with the corresponding deterministic versions \cite{NM} we can state
that the active region is reduced by the stochasticity. For example, in the
deterministic version of our second model a competition between the $C$ and
$D$ invasion processes can be observed if $9/5<b<2$ because the cooperators
invade along straight lines meanwhile the defectors win along irregular
boundaries. In this parameter range Mukherji {\it et al.} \cite{MRS}
have observed that the cooperation is eliminated when introducing stochastic
elements. This is not surprising because the $C$ invasion along straight lines
is not permitted in the stochastic models. At lower value of $b$, however,
the spatial effects can facilitate the survival of cooperators \cite{NBM,NBM2}.
In the present stochastic model the second threshold value of $b$ is
decreased by the randomness, namely, we have found $b_{c2}<1.4$
for $K=0.02$, 0.1 and 0.5\ .

The generalized mean-field approximations have clarified the importance
of short range correlations for both versions of the stochastic evolutionary
PD game inside the coexistence region. Unfortunately, this technique
is not applicable in the critical regions ($c \to 0$ and $1$) where 
long-range correlations and fluctuations play a dominant role. 

In these critical regions the MC simulations indicated clearly
power law behavior, namely $c \propto (b_{c2}-b)^{\beta}$ and
$1-c \propto (b-b_{c1})^{\beta}$) at sufficiently high noise levels.
The values of $\beta$ deduced from the MC data agree well with the 
DP exponent for both versions. These findings corroborate the
conjecture according to which the transitions in all one-component models
to an absorbing state belong to the DP universality class in the absence
of conserved symmetries. The curiosity of
the present model is that here we have two different (non-symmetric)
absorbing states whose stability regions are separated by the active phase. 
For low values of $K$ the appearance of power law behavior against $b$
is distorted by the strongly nonlinear $b$-dependence of the diffusion
and annihilation. Due to the robustness of DP universality class
similar critical behavior is expected for many other versions of
stochastic evolutionary rules.

\acknowledgements

This work was supported by the Hungarian National Research
Fund under Grant No. T-16734.

\end{document}